\documentclass[12pt]{iopart}
\usepackage{graphicx,iopams,setstack,bm}
\usepackage{subfigure}
\usepackage{multirow}

\newcommand{\be}{\begin{equation}}
\newcommand{\ee}{\end{equation}}
\newcommand{\bea}{\begin{eqnarray}}
\newcommand{\eea}{\end{eqnarray}}

\newcommand{\nn}{\nonumber}
\def\rf#1{(\ref{#1})}

\def\de#1/de#2{\frac{\partial {#1}}{\partial {#2}}}
 
\begin{document}

\title{Non minimally coupled condensate cosmologies: a phase space analysis.}

\author{ Sante Carloni$^{1}$\footnote{E-mail: sante.carloni@utf.mff.cuni.cz},\, Stefano Vignolo$^{2}$\footnote{E-mail: vignolo@diptem.unige.it} \ and
Roberto Cianci$^{2}$\footnote{E-mail: cianci@diptem.unige.it}
}
\address{ $^{1}$ Institute of Theoretical Physics, Faculty of Mathematics and Physics\\ Charles University in Prague,  V Hole\v{s}ovi\v{c}k\'{a}ch 2 180 00 Praha 8,  Czech Republic\\
$^{2}$ DIME Sez. Metodi e Modelli Matematici, Universit\`{a} di Genova\\
Piazzale Kennedy, Pad. D - 16129 Genova (Italia)}

\begin{abstract}
We present an analysis of the phase space of cosmological models based on a non minimal coupling between the geometry and a fermionic condensate. We obtain  that the strong constraint coming from the Dirac equations allows a detailed design of the cosmology of these models and at the same time guarantees an evolution towards a state indistinguishable from General Relativistic cosmological models. In this light, we show in detail  how the use of some specific potentials is able to reproduce naturally a phase of accelerated expansion. In particular we find for the first time that an exponential potential is able to induce two de Sitter phases separated by a power law expansion which could be an interesting model for the unification of an inflationary phase and a dark energy era.
\end{abstract}
\date{\today}
\maketitle

\tolerance=5000

\section{Introduction}
One of the fundamental initial hypotheses in the formulation of any relativistic theory of gravitation is the way in which matter couples to the geometry. In the first prototype of these theories, General Relativity (GR), one chooses the so called minimal coupling  i.e. matter and geometry only couple via the determinant of the metric tensor which, in the definition of the matter action, multiplies the matter Lagrangian density.

However, it is known since long time now that in the semiclassical approach to Quantum Gravity, also called quantum field theory on curved spacetimes \cite{B&D}, the appearance of non minimal couplings (NMC) is inevitable if one renormalizes the quantum stress energy tensor of the matter fields \cite{Donoghue}. NMCs also appear in the low energy limit of a number of fundamental theories in particle physics like the tree-level action of string theory \cite{strings} and in the attempt of include more naturally Mach's principle in the framework of Einstein's gravitation like in Brans-Dicke theory \cite{BD}. The study of non minimally coupled theories has accompanied the development of our understanding of Einstein's gravitational theory, but only in the last thirty years the potential importance of the role of NMCs in inflationary and dark energy models has become clear. 

In spite of these interesting features, the introduction of NMC brings also some fundamental  problems which have so far remained unsolved. For example, it is clear that,  since NMCs prescribe a different way in which these matter fields couple to gravity (in addition to the one given by the Einstein equations), the presence of NMC violates the strong equivalence principle. It has also become clear that some types of NMCs can be responsible of the appearance of ghosts and tachyons in the particle spectrum of gravitational theories \cite{Buchbinder}. Last but not least, the NMC is often modeled using a scalar field, whose nature has so far remained obscure\footnote{ Only recently, with the discovery of the Higgs field, a non minimally coupled theory of this field with the geometry has been considered \cite{Bezrukov}  which might work as a model for inflation (the minimally coupled Higgs field is known not to be a good candidate in this respect \cite{Guth}). }. 

In this paper we will explore the possibility that the scalar field typical of NMCs is not fundamental, but it is represented by a fermion condensate.   This idea was already proposed by Weinberg in the context of the Higgs field \cite{Weinberg} (see also \cite{Fabbri}) and analyzed in cosmology in \cite{Ribas}.  The motivation behind our analysis stems from the observation that, within standard Big Bang cosmology, the dynamics of the early universe has been dominated by a form of matter energy (the primordial plasma) in which fermions played an important role. At these early stages of cosmic history the effects of the quantum nature of the gravitational interaction are likely to have a non trivial role in the evolution of the Universe. Such role is modeled  by the non minimal coupling. Due to their lack of isotropy, fermions are normally  considered of little use in late time  Friedmannian cosmology. Our analysis shows that should a cosmic  background of fermions exist today as a condensate (which is not known at present), it could combine with the non minimal coupling to originate accelerated expansion. In fact we will find that in this kind of cosmologies such phase is achieved in a dynamical state which is indistinguishable from  General Relativity, thus hiding the  NMC.

We will also see that a key property of these models is given by the Dirac equation  which constraints the behavior of the condensate in such a way that it becomes straightforward to control the properties of the cosmological models. Such ``model design''  realizes naturally a feature which was for long time sought in the context of fourth order gravity \cite{Star2007}. 

The analysis we  will perform is based on the Dynamical Systems Approach (DSA). Defined by the work of Collins, Wainwright and Ellis, this technique has revealed to be a precious tool for the understanding of complex cosmological models in the framework of General Relativity \cite{Dynamical,Goliath:1998na} as well as extensions of Einstein theory \cite{SanteDynSys}.  DSA consists in recasting the cosmological equations as an autonomous system of differential equations for some tailored variables that carry also physical meaning.  Using  DSA one is able  to unfold in relative easy way a number of important aspects of  cosmological models which include the general behavior of the cosmology as well s the evolution of the shear and the occurrence of bounces. In some cases in which free functions appear in the theory the DSA has helped to select the structure of such functions \cite{Emilio}. 

In the following we will explore the phase space of cosmological models in which a non minimally coupled  fermion condensate exists and in which the fermions have a self interaction potential which depends only on the condensate itself. We will consider also an additional form of matter described by a perfect fluid.  Although the method is completely general, we will choose some simple specific forms of this potential as an example.  Among other results our analysis shows that these models, as other ``scalar tensor'' models,  can present the phenomenon of the  ``isotropization'' to General Relativity (GR) \cite{isoGR}  and at the same time show some peculiar properties, which might be used as a framework for dark energy models and/or inflation.

The paper will be divided in the following way.  Section II is a brief review of the general details of the properties of the theory we will consider in this paper and its key equations in the Friedmann-Lema\^{\i}tre-Robertson-Walker (FLRW) metric. Some exact results in vacuum are also presented. Section III deals with the dynamical systems analysis in presence and absence of a perfect fluid and for different potentials. Finally Section IV gives the conclusions.

Unless otherwise specified, natural units ($\hbar=c=k_{B}=8\pi G=1$) will be used throughout this paper, Latin and Greek indices run from 0 to 3. The symbol $\nabla$ represents the Levi--Civita covariant derivative associated with a metric tensor $g_{ij}$. We use the $+,-,-,-$ signature for the metric tensor and the Riemann tensor is defined by
\begin{equation}
R^{d}_{\;\;cab}
=\partial_a\Gamma_{bc}^{\;\;\;d} - \partial_b\Gamma_{ac}^{\;\;\;d} +
\Gamma_{ap}^{\;\;\;d}\Gamma_{bc}^{\;\;\;p}-\Gamma_{bp}^{\;\;\;h}\Gamma_{ac}^{\;\;\;p}\;,
\end{equation}
where the $\Gamma_{ab}^{\;\;\;c}$ are the Christoffel symbols associated with the metric $g_{ij}$, defined by
\begin{equation}
\nabla_{\partial_a}\partial_b = \Gamma_{ab}^{\;\;\;c}\,\partial_{c}\,.
\end{equation}
The Ricci tensor is obtained by contracting the {\em first} and the {\em third} index via the metric $g_{ab}$
\begin{equation}\label{Ricci}
R_{ab}=R^{c}_{\phantom{c}acb}\,.
\end{equation}
\section{The $(1+\epsilon\;\bar\psi\psi)R$-theory in the FLRW metric.}
Let us consider an action of the form (also proposed in a more complicated form in \cite{Ribas,Souza})
\be\label{1.1}
{\cal S}= \int{\sqrt{|g|}\left[\left(1 + \epsilon\bar\psi\psi\right)\/R - L_D\right]ds},
\ee
where the Einstein--Hilbert term is non--minimally coupled to the condensate $\bar\psi\psi$ of a Dirac field whose  Lagrangian has the form
\be\label{2.1}
L_D = + \frac{i}{2}\left( \bar{\psi}\Gamma^iD_i\psi - D_i\bar{\psi}\Gamma^i\psi\right) -m\bar{\psi}\psi + V(\bar\psi\psi),
\ee
where a fermionic self--interaction potential $V(\bar\psi\psi)\/$ is present. In eq. \rf{1.1} $\epsilon$ indicates a suitable constant parameter, while in eq. \rf{2.1} we have $\Gamma^i :=e^i_\mu\gamma^\mu\/$ ($\gamma^\mu\/$ representing Dirac matrices and $e^\mu_i\/$ a tetrad field such that $g_{ij}=e^\mu_ie^\nu_j\eta_{\mu\nu}$) and  the $D_i\/$ indicate covariant derivatives of the spinor field 
\begin{eqnarray}
&&	D_i\psi = \partial_i\psi - \Omega_i\psi,\label{3.1a}\\
&&	D_i\bar{\psi} = \partial_i\bar{\psi} + \bar{\psi}\Omega_i,\label{3.1b}
\end{eqnarray}
where
\begin{equation}\label{4.1}
\Omega_i = -\frac{1}{4}g_{ij}\left(\Gamma^{\;\;\;j}_{pq} - e^j_\mu\partial_p e^\mu_q\right)\Gamma^p\Gamma^q.
\end{equation}
Introducing for simplicity the notation $\varphi := \bar\psi\psi$, it is easily seen that the action \rf{1.1} yields Einstein--like field equations of the form
\begin{equation}\label{5.1}
\left(1 + \epsilon\varphi\right)\left(R_{ij} - \frac{1}{2}Rg_{ij}\right)= \Sigma_{ij} + \epsilon\left(\nabla_i\nabla_j\/\varphi - g_{ij}g^{pq}\nabla_p\nabla_q\/\varphi\right),
\end{equation} 
where 
\be\label{6.1}
\Sigma_{ij} = + \frac{i}{4}\left( \bar{\psi}\Gamma_{(i}D_{j)}\psi - D_{(i}\bar{\psi}\Gamma_{j)}\psi\right) - \frac{1}{2}L_D\,g_{ij},
\ee
is the energy--momentum tensor of the Dirac field, obtained by variation of the Dirac Lagrangian $L_D$ with respect to the tetrad field. At the same time, from the action \rf{1.1} we derive Dirac equations for the spinor field of the form 
\begin{eqnarray}
&& i\Gamma^iD_i\psi -m\psi + V'(\varphi)\psi - \epsilon\psi\/R =0,\label{7.1a}\\
 && iD_i\bar{\psi}\Gamma^i + m\bar\psi - V'(\varphi)\bar\psi + \epsilon\bar\psi\/R =0,\label{7.1b}
\end{eqnarray}
where $V' = \frac{dV}{d\varphi}$. Making use of eqs. (\ref{7.1a}-\ref{7.1b}) we can express the energy--momentum tensor \rf{6.1} as
\be\label{8.1}
\fl \Sigma_{ij} = + \frac{i}{4}\left( \bar{\psi}\Gamma_{(i}D_{j)}\psi - D_{(i}\bar{\psi}\Gamma_{j)}\psi\right) - \frac{\epsilon}{2}\varphi\/R\,g_{ij} -\frac{1}{2}V(\varphi)\,g_{ij} + \frac{1}{2}\varphi\/V'(\varphi)\,g_{ij}.
\ee 
Now, let us consider a spatially flat Friedmann-Lema\^{\i}tre-Robertson-Walker (FLRW) metric tensor
\begin{equation}\label{9.1}
ds^2 = dt^2 - a(t)^2\/\left( dx^2 + dy^2 + dz^2 \right).
\end{equation}
The tetrad field associated with the metric \rf{9.1} is expressed as
\begin{equation}\label{10.1}
e^\mu_0 = \delta^\mu_0, \qquad e^\mu_A = a(t)\delta^\mu_A, \qquad A=1,2,3. 
\end{equation}
From this, it is easily seen that the $\Gamma^i =e^i_\mu\gamma^\mu\/$ matrices are given by
\begin{equation}\label{11.1}
\Gamma^0 = \gamma^0, \quad \Gamma^A = \frac{1}{a(t)}\delta^A_\mu\gamma^\mu.
\end{equation}
as well as that the coefficients of the spin connection are 
\begin{equation}\label{12.1}
\Omega_0 =0, \qquad \Omega_A = \frac{\dot a}{2}\gamma^A\gamma^0, \qquad A=1,2,3.
\end{equation}
Due to eqs. \rf{9.1}--\rf{12.1}, in the metric \rf{9.1}  the Dirac equations (\ref{7.1a}-\ref{7.1b}) assume the 
form
\begin{eqnarray}
&&	\dot\psi + \frac{3}{2}\frac{\dot a}{a}\psi + im\gamma^0\psi - V'(\varphi)\gamma^0\psi + i\epsilon\/R\gamma^0\psi =0,\label{13.1a}\\
&&	\dot{\bar\psi} + \frac{3}{2}\frac{\dot a}{a}\bar\psi - im\bar{\psi}\gamma^0 + V'(\varphi)\bar\psi\gamma^0 - i\epsilon\/R\bar\psi\gamma^0 =0,\label{13.1b}
\end{eqnarray}
From eqs. (\ref{13.1a}-\ref{13.1b}) we derive the evolution law for the scalar field $\varphi =\bar\psi\psi$
\be\label{14.1}
\dot\varphi + 3\frac{\dot a}{a}\varphi =0,
\ee
thus obtaining the final relation
\be\label{18.1}
\varphi =\frac{\varphi_0}{a^3}.
\ee
This result is independent of the form of the gravitational action and  constitutes a very tight constraint on the entire theory. Since one knows that $\varphi\rightarrow0$ when the scale factor grows, the non-minimal coupling can be used as a ``switch''  at the action level to regulate the onset  of the different terms of the Lagrangian. This is not possible in standard scalar tensor theories because in that case the behavior of the scalar field is described by a Klein-Gordon equation, whose solution is in general much more complicated.

Making use again of eqs. (\ref{9.1}-\ref{13.1b}), it is a straightforward matter to verify that the non vanishing components of the fermionic energy--momentum tensor \rf{8.1} are represented by
\begin{eqnarray}
&&	(\Sigma_D)_{00} = +\frac{m}{2}\varphi - \frac{1}{2}V(\varphi),\label{15.1a}\\
&&	(\Sigma_D)_{AA} = + \frac{\epsilon}{2}\varphi\/R\/a^2 + \frac{1}{2}V(\varphi)\/a^2 - \frac{1}{2}\varphi\/V'(\varphi)\/a^2  \qquad A=1,2,3. \label{15.1b}
\end{eqnarray}
Inserting the content of eqs. (\ref{15.1a}-\ref{15.1b}) into eqs. \rf{5.1}, the Einstein--like equations assume the following expression
\begin{eqnarray}
&&
(1+\epsilon\varphi)\/3\left(\frac{\dot a}{a}\right)^2 = \frac{m}{2}\varphi -3\epsilon\frac{\dot a}{a}{\dot\varphi} - \frac{1}{2}V(\varphi),\label{16.1a}\\
&& (1+\epsilon\varphi)\left[2\frac{\ddot a}{a} + \left( \frac{\dot a}{a} \right)^2\right] = -\frac{\epsilon}{2}\varphi\/R - \epsilon\ddot\varphi - 2\epsilon\frac{\dot a}{a}\dot\varphi -\frac{1}{2}V(\varphi) +\frac{1}{2}\varphi\/V'(\varphi).\label{16.1b}
\end{eqnarray}
We can replace eq. \rf{16.1b} by the equivalent Raychaudhuri equation
\be\label{17.1}
(1+\epsilon\varphi)6\frac{\ddot a}{a} = -\frac{3}{2}\epsilon\varphi\/R - 3\epsilon\ddot\varphi - 3\epsilon\frac{\dot a}{a}\dot\varphi - \frac{m}{2}\varphi - V(\varphi) + \frac{3}{2}\varphi\/V'(\varphi),
\ee

In the absence of the self--interaction potential ($V(\varphi)=0$), inserting \rf{18.1} into \rf{16.1a} we obtain the final differential equation
\be\label{19.1}
{\dot a}^2 = \frac{m}{2}{\left[ \frac{3a}{\varphi_0} - \frac{6\epsilon}{a^2} \right]}^{-1}.
\ee
For values $a<<1$ eq. \rf{19.1} can be approximated to 
\be\label{20.1}
{\dot a}^2 = - \frac{m\/a^2}{12\epsilon},
\ee
this solution for $\epsilon<0$ admits a de Sitter solution $a(t)= a_0\exp(\lambda t)$ with $\lambda := \sqrt{\frac{m}{-12\epsilon}}$ that can  describe inflationary models or dark energy eras. Instead,  for $\epsilon>0$ the real part of the  solution of \rf{20.1} becomes oscillatory.  In the case $a >>1$ eq. \rf{19.1} can be approximated to
\be\label{21.1}
{\dot a}^2 = \frac{m\varphi_0}{6a},
\ee
which possesses a Friedmann solution $a(t)={\left[\frac{3}{2}\left(\lambda\/t + c\right)\right]}^{\frac{2}{3}}$ with $\lambda:=\sqrt{\frac{m\varphi_0}{6}}$. This means that in this case the theory can describe a transition from a dark energy era (a period characterized by accelerated expansion)  to a Friedmann one (a period characterized by a decelerated expansion).

In order to add a perfect fluid to our cosmological model, we suppose a barotropic perfect fluid assigned, with equation of state $p=w\rho$ ($w\in [0,1[$) and standard conservation law
\be\label{1.3}
\dot\rho + 3\frac{\dot a}{a}\left(\rho + p\right)=0,
\ee
yielding the relation
\be\label{4.3}
\rho=\frac{\rho_0}{a^{3(1+w)}}.
\ee
The field equations (\ref{16.1a} -\ref{16.1b}) and \rf{17.1} become 
\begin{eqnarray}
&&\fl(1+\epsilon\varphi)\/3\left(\frac{\dot a}{a}\right)^2 = \rho + \frac{m}{2}\varphi - \frac{1}{2}V(\varphi) -3\epsilon\frac{\dot a}{a}{\dot\varphi},\label{2.3a}\\
&&
\fl(1+\epsilon\varphi)\left[2\frac{\ddot a}{a} + \left( \frac{\dot a}{a} \right)^2\right] = -p -\frac{\epsilon}{2}\varphi\/R -\frac{1}{2}V(\varphi) +\frac{1}{2}\varphi\/V'(\varphi) - \epsilon\ddot\varphi - 2\epsilon\frac{\dot a}{a}\dot\varphi,\label{2.3b}
\end{eqnarray}
and
\be\label{3.3}
\fl(1+\epsilon\varphi)6\frac{\ddot a}{a} = -(\rho + 3p) -\frac{3}{2}\epsilon\varphi\/R - 3\epsilon\ddot\varphi - 3\epsilon\frac{\dot a}{a}\dot\varphi - \frac{m}{2}\varphi -V(\varphi) + \frac{3}{2}\varphi\/V'(\varphi).
\ee
The complete system of equations  is therefore
\begin{equation}
3(1-2\epsilon\varphi)\left(\frac{\dot a}{a}\right)^2 = \rho + \frac{m}{2}\varphi - \frac{1}{2}V(\varphi),\label{FriedDSA}
\end{equation}
\begin{equation}
6(1-2\epsilon\varphi)\frac{\ddot a}{a} = -(\rho + 3p) - 18\epsilon\varphi\left(\frac{\dot a}{a}\right)^2 - \frac{m}{2}\varphi -V(\varphi) + \frac{3}{2}\varphi\/V'(\varphi),
\end{equation}
\begin{equation}
\dot\varphi + 3\frac{\dot a}{a}\varphi =0,
\end{equation}
\begin{equation}
\dot\rho + 3\frac{\dot a}{a}\left( 1+ w\right)\rho=0,\label{ConsMatt}
\end{equation}
where we have already substituted the evolution equation for the scalar field \rf{14.1} and the expression of the Ricci scalar in flat FLRW spacetime $R=-6\left(\frac{\ddot a}{a}+\frac{\dot{a}^2}{a^2} \right)$. These equations will be the starting point of our analysis.
\section{Dynamical Systems Analysis}
Let us now analyze the cosmology deriving from eqs. (\rf{FriedDSA}-\rf{ConsMatt}), using the Dynamical Systems Approach. This method consists in rewriting the cosmological equations written above in terms of some dimensionless variables (including the time variable), that play the same role of  the $\Omega$ parameters in standard Friedmannian cosmology and therefore carry a precise physical meaning. 

Choosing the variables
\begin{equation} \label{Var}
\fl\Omega =\frac{\rho}{3 H^2 (1-2 \epsilon  \varphi)},\quad X=\frac{\epsilon  \varphi}{1-2 \epsilon  \varphi},\quad Y=\frac{V(\varphi)}{6 H^2 (1-2 \epsilon  \varphi)}, \quad M=\frac{m\varphi }{6 H^2 (1-2 \epsilon  \varphi )},
\end{equation}
 where $H=\frac{\dot a}{a}$ and the logarithmic time  ${\mathcal N}=\ln a$. In terms of $X Y,\Omega, M$, the cosmological equations can be written as
\begin{eqnarray}
&&\frac{dX({\mathcal N})}{d{\mathcal N}}= -3X (1+2X),\label{DynSys1}\\
&&\frac{dY({\mathcal N})}{d{\mathcal N}}= Y
   \left[M+(3 w+1) \Omega -3 \mathcal{V}+2\right]+Y^2 \left[2-3 \mathcal{V}\right],\label{DynSys2}\\
&&\frac{dM({\mathcal N})}{d{\mathcal N}}= M\left[M+(3 w+1) \Omega +Y \left(2-3
   \mathcal{V}\right)-1\right],\label{DynSys3}\\
&& \frac{d\Omega({\mathcal N})}{d{\mathcal N}}=\Omega  \left[M-3 w-3 Y \mathcal{V}+2 Y-1+(3 w+1)\Omega\right] ,\label{DynSys4}
\end{eqnarray}
with the constraint 
\begin{equation}
 1-M+Y-\Omega=0,\label{constraint}
\end{equation}   
 coming from \rf{FriedDSA}. In the above equations the characteristic function $\mathcal{V}$
\begin{equation}
 \mathcal{V} =\mathcal{V}\left(\frac{X}{2 X \epsilon +\epsilon }\right)= \left.\frac{\varphi V'(\varphi)}{V(\varphi)}\right|_{\varphi=\varphi(X,\epsilon )},
\end{equation}
contains the information on the form of the potential $V(\varphi)$ in the (\rf{FriedDSA}-\rf{ConsMatt}).

Note that the  variable $X$ is zero for $\varphi\rightarrow0$,  $X=-1/2$ for $\varphi\rightarrow\infty$ and $X\rightarrow \pm\infty$ when $\varphi\rightarrow (1/2\epsilon)^\pm$. Thus the ``X direction'' represents the complete evolution of the field $ \varphi$ and while the fixed points with $X$ coordinate $-1/2$ will be unphysical the ones in $X=0$ will represent effectively GR. In this way, every time there is an attractor on the  $X=0$ axis it is a signal that the model converges to GR. This is a typical feature of theories which have a ``scalar tensor'' structure.  In addition, since \rf{18.1} holds we also have that $X=-1/2$ for $a\rightarrow0$ i.e. ``early time''  and $X=0$ for $a\rightarrow\infty$ i.e. ``late time'', so the fixed points on these invariant submanifolds will characterize early and late time solutions for the theory\footnote{It is clear that such definition holds only when the solution for the scale factor has a monotonic character which is not obvious in NMC cosmologies. However  the orbits plotted in the phase space represent only evolutions in which  $a$ is monotonic. Would $a$ change its character, its derivative would be zero and consequently $H$ would be zero, but this can happen only asymptotically. In this respect then the above definition of  ``early time'' and ``late time'' can be used without confusion.}. 

The above system  admits five invariant submanifolds  ($X=0$, $X=-1/2$, $Y=0$, $M=0$, $\Omega=0$), so there can be no global attractor in these cosmologies. However,  the origin of the coordinate axes can be an attractor for a large set of initial conditions. This also means that for these models GR can be an attractor for the same set of these conditions. Such feature is not at all common in scalar tensor gravity: usually the set of initial conditions is usually much smaller \cite{Jannie}. 

The solutions associated with the fixed points of the above system when $\beta\neq0$ can be found using the equations
\begin{eqnarray}
&& a=a_0(t-t_0)^{1/\beta}, \\
&& \rho=\rho_0(t-t_0)^{-3(1+w)/\beta},\\
&& \varphi=\varphi_0(t-t_0)^{-3/\beta},\\
&& 2 \beta=2+M_0+6 X_0-(3 \mathcal{V}_0 -2) Y_0+(1+3 w)\Omega_0, \label{Alpha}
\end{eqnarray}
where the subscript ``0'' refers to the value of the variable in the fixed point. In the following will make some illustrative choices  for $V$ and as consequence for $\mathcal{V}$.
\subsection{Dynamical Systems Analysis in absence of perfect fluid(s).}
We analyze first the  case of absence of perfect fluid ($\Omega=0$) so that the dynamical above system  loses an equation. Implementing \rf{constraint} to eliminate $Y$, one obtains
\begin{eqnarray}
&&\frac{dX({\mathcal N})}{d{\mathcal N}}= -3X (1+2X),\label{DynSys1NOm}\\
&&\frac{dM({\mathcal N})}{d{\mathcal N}}= M^2+M \left[(M-1) \left(2-3
   \mathcal{V}\right)-1\right],\label{DynSys2NOm}\\
&& Y=M-1.
\end{eqnarray}
The phase space has dimension two and can be easily plotted. We will analyze this system in two cases:  $V(\varphi)=V_0 \varphi^\alpha$ and  $V(\varphi)=V_0 \exp(-\lambda \varphi)$.

\subsubsection{The case $V(\varphi)=V_0 \varphi^\alpha$ }
As a first example, let us consider the potential $V(\varphi)=V_0 \varphi^\alpha$. This type of potential is the most common in the treatment of interacting fermions \cite{NJL}. Because of the \rf{18.1} at early times ($a\rightarrow0$) the scalar field will have high values and, depending on the sign  of the parameter $\alpha$, the potential will be negligible or dominant.  The converse happens  at late time at late times ($a\rightarrow\infty$). This allows a certain degree of control on the cosmological model.

In terms of the above variables, the considered potential is  characterized by $\mathcal{V}=\alpha$ and we have
\begin{eqnarray}
&&\frac{dX({\mathcal N})}{d{\mathcal N}}=-3X (1+2X),\label{DynSys1NOm}\\
&&\frac{dM({\mathcal N})}{d{\mathcal N}}= M^2+M \left[(M-1)  \left(2-3
   \alpha\right)-1\right].\label{DynSys2NOm}
\end{eqnarray}
This dynamical system has four fixed points shown in Table \ref{TabNOm}. Their coordinates do not depend on the parameters of the system and therefore we can expect them to be present also in other cases together with other fixed points. The solutions associated with the fixed points correspond to power law behaviors for the scale factor, whose character depends on the choice of $\alpha$.  The character of the scale factor depends on $\alpha$ for the points $\mathcal{A}$  and $\mathcal{C}$. For $\alpha<0$ both these points always represent a contraction.  For $0<\alpha<2/3$,  $\mathcal{A}$  represents a contraction\footnote{The fact that this point is associate to a contraction might appear in contrast with the statement made above in which along an orbit $H$ does not change sign. However, looking at the stability one can see that for $\alpha<1$,  $\mathcal{A}$ is a saddle and therefore an orbit approaching to is simply implies a slowing of the expansion rate which magnitude depends on the distance between the orbit and the fixed point.} and $\mathcal{C}$ an accelerated expansion. For $2/3<\alpha<1$,  $\mathcal{A}$  represents a contraction and $\mathcal{C}$ a decelerated expansion. For $1<\alpha<5/3$, $\mathcal{A}$  represents an accelerated expansion (power law inflation) and $\mathcal{C}$  a decelerated expansion. Finally for $\alpha>5/3$, $\mathcal{A}$ represents a decelerated expansion and $\mathcal{C}$  a decelerated one.

Note that only two  ($\mathcal{C}$, $\mathcal{D}$) of the fixed points in  Table  \ref{TabNOm}  can represent physical solutions of the cosmological equations  i.e. they correspond to actual solutions of the cosmological equations. In particular, this happens for $\alpha<1$ for $\mathcal{C}$ and for $\alpha>1$ for $\mathcal{D}$. For the other points, it is easy to see that these behaviors for the scale factor can be obtained by the \rf{FriedDSA} in the right approximation. In the case of point $\mathcal{A}$ the solution $a=a_0\left(t-t_0\right)^{\frac{2}{3(\alpha-1)}}$  can be obtained from the \rf{FriedDSA}  in the approximation $\varphi\gg1$ and $m\approx0$, which is suggested by its coordinates. The same happens for the solution associated with the point $\mathcal{B}$: it can be derived assuming  $\varphi\gg1$ and $V(\varphi)\ll 1$. Like for the physical points, one has that the approximated solutions are consistent with the theory  only for specific intervals of the parameter  $\alpha$. For example, for the point  $\mathcal{B}$ the conditions $\varphi\gg1$ and $V(\varphi)\ll 1$ are consistent only if $\alpha<1$.
\begin{table}[tbp] \centering
\caption{The fixed points and the solutions of the purely fermion NMC model with $V(\varphi)=V_0 \varphi^\alpha$. }
\begin{tabular}{cccccccc}
& & \\
\hline   Point &$(X,M,Y)$ &  Scale Factor & Condensate \\ \hline
& & \\
$\mathcal{A}$ &$\left(-\frac{1}{2},0,-1\right)$&  $a=a_0\left(t-t_0\right)^{\frac{2}{3(\alpha-1)}}$ & $\varphi\rightarrow\infty
$\\ 
& & \\
$\mathcal{B}$& $\left(-\frac{1}{2},1,0\right)$& $a=a_0 \exp\left(\sqrt{-\frac{m}{12\epsilon}}(t-t_0)\right)$&$\varphi\rightarrow\infty$  \\
& & \\
$\mathcal{C}$  &$\left(0,0,-1\right)$&  $a=a_0\left(t-t_0\right)^{\frac{2}{3\alpha}}$ & $\varphi=0$\\ 
& & \\
$\mathcal{D}$& $\left(0,1,0\right)$& $a=a_0\left(t-t_0\right)^{2/3}$&$\varphi=0
$ \\
& & \\ \hline
\end{tabular}\label{TabNOm}
\end{table}

The stability of these fixed points can be found using the  Hartman-Grobman (HG) theorem and it is given together with their coordinates in Table \ref{TableStabNOm}.  Note that the critical values for the change in stability coincide with the ones related to the above mentioned issues of the consistency of the solutions. Of all the fixed points only $\mathcal{C}$ and $\mathcal{D}$ can be attractors. Since both the fixed points lay on $X=0$ invariant submanifold and this submanifold represents states indistinguishable from GR, the fact that these points are attractors implies there exist a set of initial conditions for which the theory essentially evolves towards GR.  These initial conditions are
given by
\begin{eqnarray}
&&X_0>-\frac{1}{2}, \quad M_0<1, \quad \alpha<1,\\
&&X_0>-\frac{1}{2}, \quad M_0>0, \quad \alpha>1.
\end{eqnarray}
\begin{table}[tbp] \centering
\caption{The fixed points and their stability of the purely fermion NMC model with $V(\varphi)=V_0 \varphi^\alpha$. Here R= repeller, A=attractor and S=saddle point.}
\begin{tabular}{cccccccc}
& & \\
\hline   Point &$(X,M,Y)$ & Eigenvalues &  Stability $\alpha>1$ & Stability $\alpha<1$ \\ \hline
& & \\
$\mathcal{A}$ &$\left(-\frac{1}{2},0,-1\right)$&$\left[3, 3 (\alpha -1)\right]$ &  R&  S\\ 
& & \\
$\mathcal{B}$& $\left(-\frac{1}{2},1,0\right)$&$\left[3, 3 (1-\alpha )\right]$& S&  R \\
& & \\
$\mathcal{C}$ &$\left(0,0,-1\right)$& $\left[-3, 3 (\alpha -1)\right]$&S&  A\\ 
& & \\
$\mathcal{D}$& $\left(0,1,0\right)$& $\left[-3, 3 (1-\alpha )\right]$&A&  S \\
& & \\ \hline
\end{tabular}\label{TableStabNOm}
\end{table}

\begin{figure}[htbp]
\includegraphics[scale=0.7]{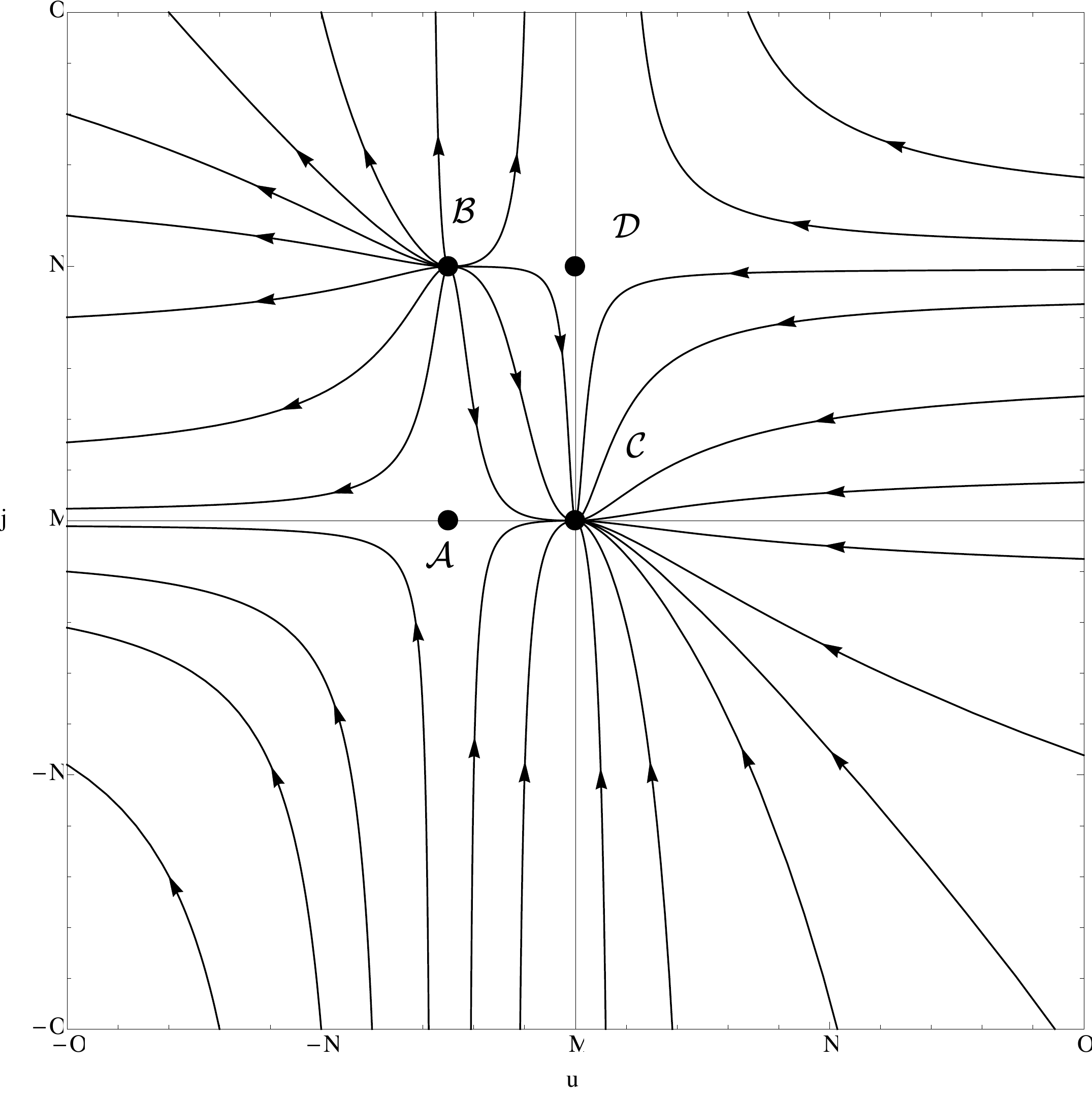}
\centering
\caption{Phase space of the purely fermion NMC model with $V(\varphi)=V_0 \varphi^\alpha$ for $\alpha<1$.}
\label{PlotNOm1}
\end{figure}

\begin{figure}[htbp]
\includegraphics[scale=0.7]{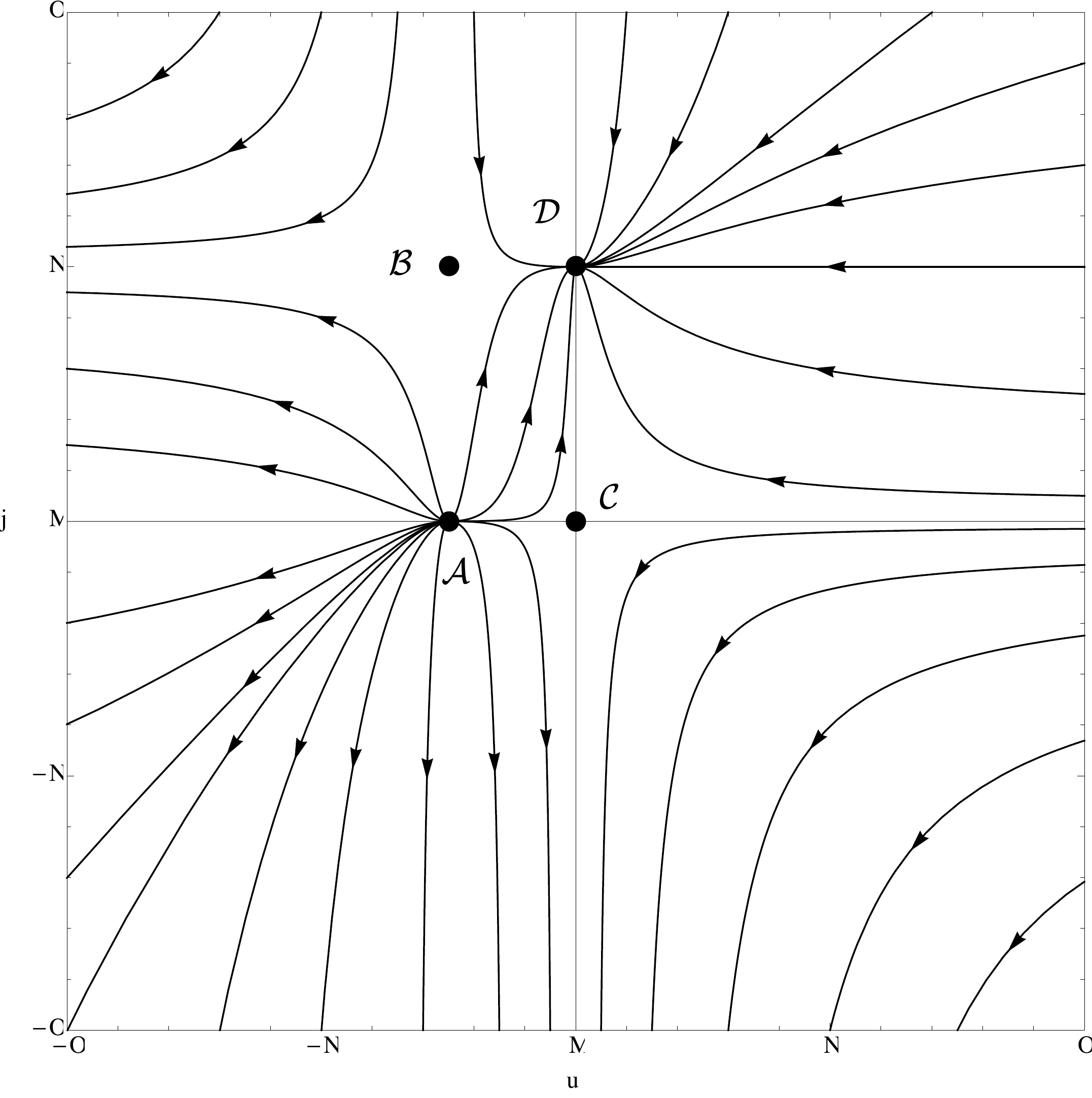}
\centering
\caption{Phase space of the purely fermion NMC model with $V(\varphi)=V_0 \varphi^\alpha$ for $\alpha>1$.} 
\label{PlotNOm2}
\end{figure}

Some examples of phase space are  plotted in Figures \ref{PlotNOm1} and \ref{PlotNOm2}.


\subsubsection{The case $V(\varphi)=V_0 \exp(-\lambda \varphi)$ }
Our interest in this type of potential is due to the fact that in the limit $\varphi\rightarrow0$ it becomes a cosmological constant term, introducing in this way a dynamical realization of the cosmological constant related to the condensate. In fact, since eq. \rf{18.1} holds, we know that at early time the potential will be irrelevant, reducing the theory to the case analyzed in \rf{19.1}, whereas at late time the potential becomes effectively a constant. Let us see how this behavior is realized in terms of the phase space.

For $V(\varphi)=V_0 \exp(-\lambda \varphi)$ we have $\mathcal{V}=-\frac{\lambda  X}{(2 X+1) \epsilon }$  and the associated dynamical system becomes
\begin{eqnarray}
&&\frac{dX({\mathcal N})}{d{\mathcal N}}= -3X (1+2X),\label{DynSys1NOmEx}\\
&&\frac{dM({\mathcal N})}{d{\mathcal N}}= 3M(M-1) \left(1+\frac{ \lambda  X}{\epsilon(1+2 X )}\right),\label{DynSys2NOmEx}\\
&& Y=M-1.\label{ConstrEx}
\end{eqnarray}
Note that in this case the above system  presents a singularity in the line $X=-1/2$. In the variables we have used, this is in fact the most common case. The singularity however is purely a result of our parametrization and has no correspondence in the actual equations. Therefore we can try to work around the singularity using a change of coordinates.

Setting for example\footnote{One might be tempted to redefine the dimensionless time in this way and  treat the entire system in this way. However, in general the sign of the quantity $1-2\epsilon \varphi$ can be subject to a change and this might lead to problem in the interpretation of the orientation of the flow. One could of course use  the factor $\left(1-2\epsilon \varphi\right)^2$ to ensure the monotonicity of the time coordinate. Such choice would lead to a system in which a fixed sub manifold  ($X=1/2$) would appear. This would not change the topology of the flow, but it would complicate the dynamical system analysis. For this reason we will keep our original choice using the analysis of the original system as a control tool.}  $d{\mathcal M}=d{\mathcal N} (1-2\epsilon \varphi)$  one obtains
\begin{eqnarray}
&&\frac{dX({\mathcal M})}{d{\mathcal M}}= -3 X(1+ 2X)^2\label{DynSys1NOmExReg}\\
&&\frac{dM({\mathcal M})}{d{\mathcal M}}= \frac{3}{\varepsilon}M (M-1)  \left[\lambda  X+\epsilon (1+2 X )\right]\label{DynSys2NOmExReg}\\
&& Y=M-1.
\end{eqnarray}
The system (\ref{DynSys1NOmEx}-\ref{ConstrEx})  has four fixed points shown in Table \ref{TabNOmEx}. 

Note that for the point $\mathcal{A}$ the exponent \rf{Alpha} is divergent. In fact, looking at the variable definitions, we can see that this point corresponds to an inconsistent relation between the quantities in the cosmological equations. As we will see the point is always unstable and therefore it poses no issue for understanding the dynamics of the cosmology. For the other solutions, the same type of reasoning given in the previous section holds: we have two points which are unphysical ($\mathcal{A}$, $\mathcal{B}$) and two which are physical ($\mathcal{C}$, $\mathcal{D}$). The difference is that the parameter $\lambda$ does not have the same weight of the parameter $\alpha$ in the previous case. This can be understood thinking that the change in the power law potential due to a change in $\alpha$ is a more radical modification of the change of the ``time constant'' of the exponential potential.  Note also that the solutions associated with the fixed points are the same except the one associated with the point $\mathcal{C}$. This result is expected since this point is the only one that expresses a potential dominated solution.
\begin{table}[tbp] \centering
\caption{The fixed points and the solutions of the purely fermion NMC model with $V(\varphi)=V_0 \exp(-\lambda \varphi)$.}
\begin{tabular}{cccccccc}
& & \\
\hline   Point &$(X,M,Y)$ &  Scale Factor & Condensate \\ \hline
& & \\
$\mathcal{A}$ &$\left(-\frac{1}{2},0,-1\right)$&  N/A& N/A\\ 
& & \\
$\mathcal{B}$& $\left(-\frac{1}{2},1,0\right)$& $a=a_0 \exp\left(\sqrt{-\frac{m}{12\epsilon }}(t-t_0)\right)$&$\varphi\rightarrow\infty $  \\
& & \\
$\mathcal{C}$  &$\left(0,0,-1\right)$&  $a=a_0 \exp\left(\sqrt{-\frac{V_0}{6}}(t-t_0)\right)$ & $\varphi=0$\\ 
& & \\
$\mathcal{D}$& $\left(0,1,0\right)$& $a=a_0\left(t-t_0\right)^{2/3}$&$\varphi=0$ \\
& & \\ \hline
\end{tabular}\label{TabNOmEx}
\end{table}

The stability of these fixed points can be found using the HG theorem and it is given together with their coordinates in Table \ref{TableStabNOmEx}. As expected  from the considerations on the potential, the point $\mathcal{C}$ (the potential dominated fixed point) is always an attractor. In addition, since the points $\mathcal{A}$ and $\mathcal{B}$ have one zero eigenvalue, their stability has to be  determined using the Central Manifold Theorem (CMT) \cite{Car, Abdelwahab:2007jp}. 

\begin{figure}[htbp]
\includegraphics[scale=0.50]{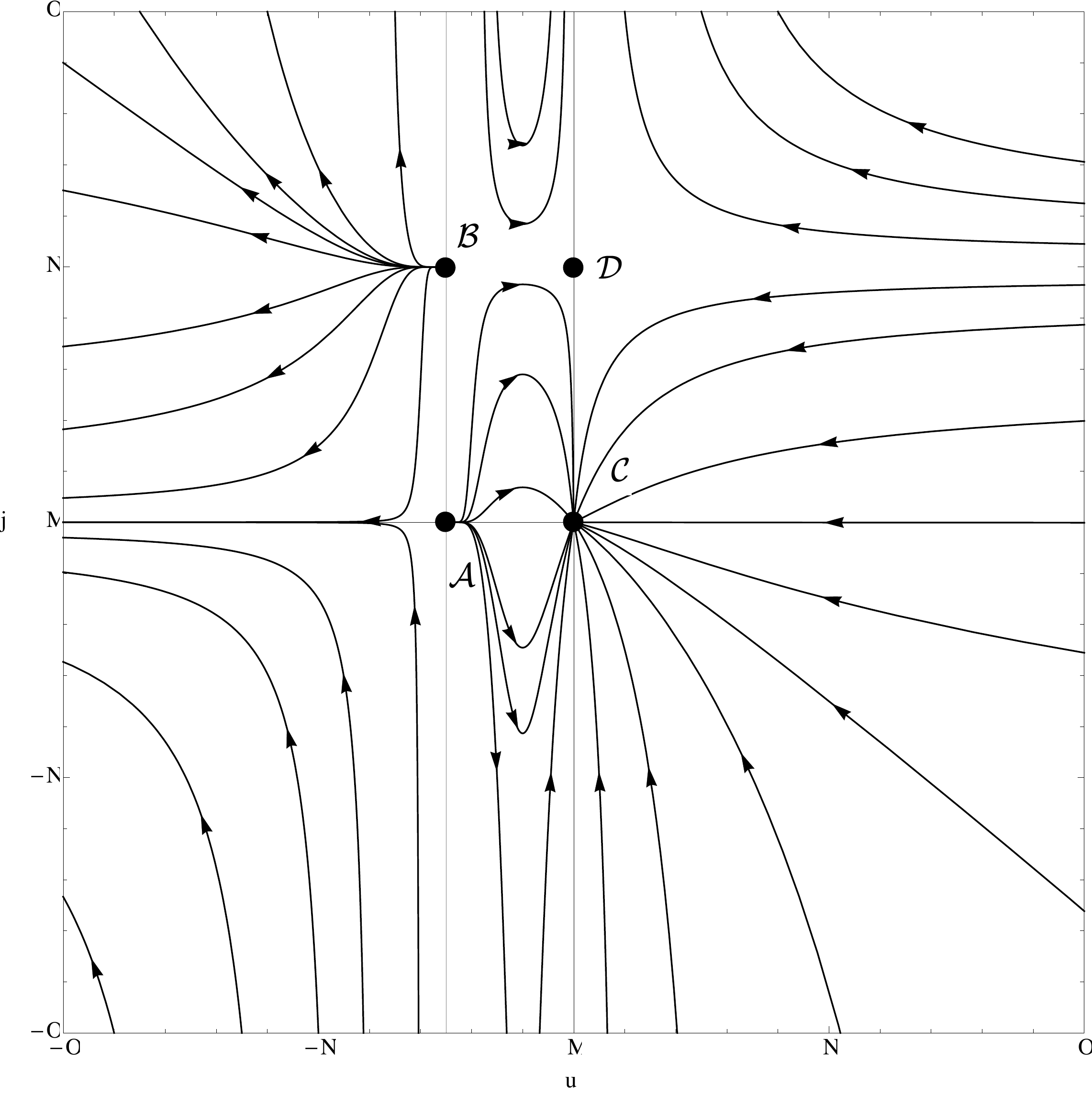}
\centering
\caption{Phase space of the purely fermion NMC model with $V(\varphi)=V_0 \exp(-\lambda \varphi)$ for $\epsilon>0$. }
\label{PSExp01}
\end{figure}

\begin{figure}[htbp]
\includegraphics[scale=0.50]{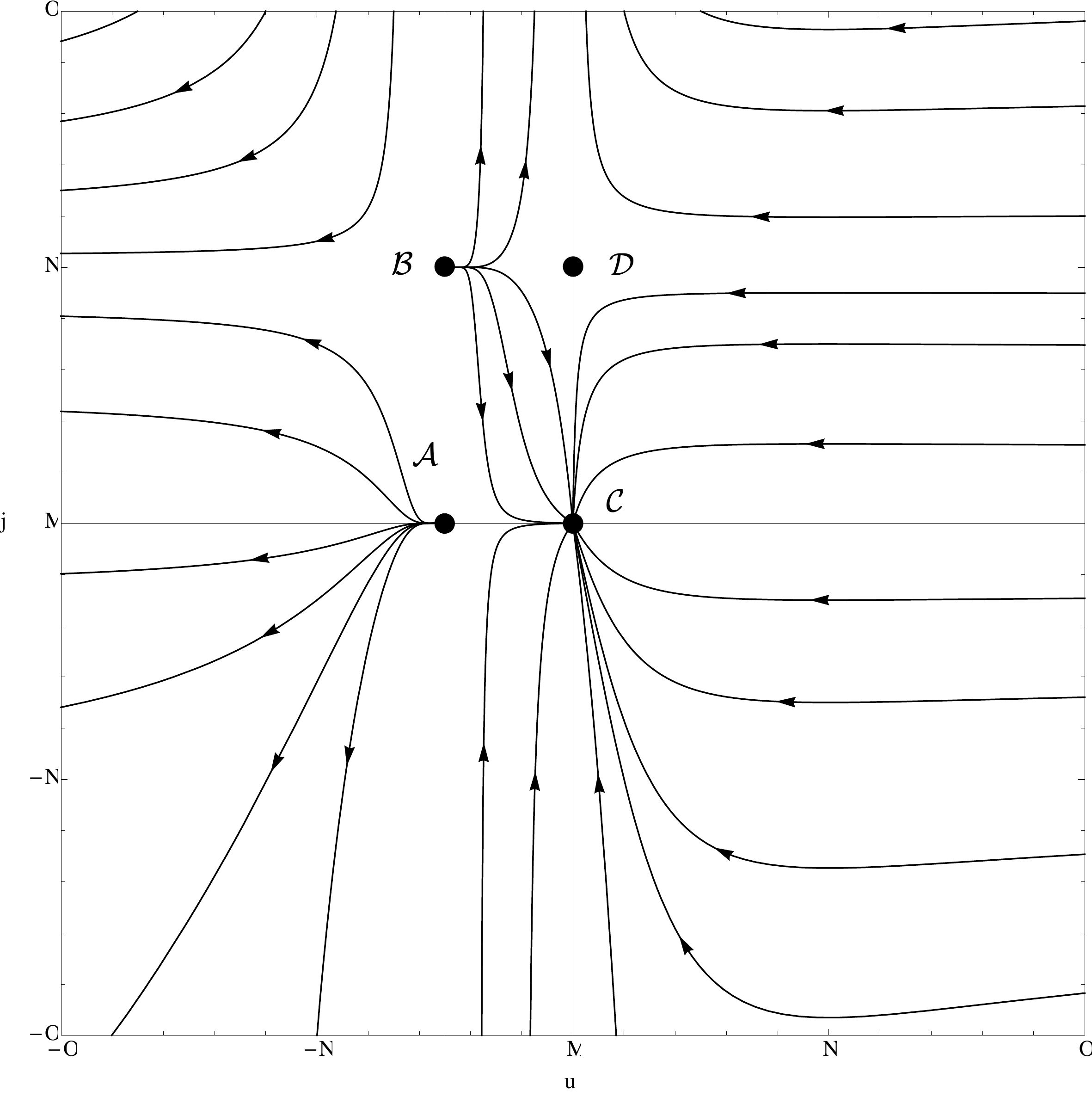}
\centering
\caption{Phase space of the purely fermion NMC model  with $V(\varphi)=V_0 \exp(-\lambda \varphi)$ for $\epsilon <0$.}
\label{PSExp02}
\end{figure}

Let us consider for example the point $\mathcal{A}$.  One can write the system (\ref{DynSys1NOmExReg}-\ref{DynSys2NOmExReg})  in the form
\begin{eqnarray}
&&\frac{dX({\mathcal M})}{d{\mathcal M}}= C X  +F(X, M)\\
&&\frac{dM({\mathcal M})}{d{\mathcal M}}= PM+G(X,M)
\end{eqnarray}
where $C$ and $P$ correspond to the linear part of the equation and $F$ and $G$ to the non--linear one.  The CMT says that the behavior of  the fixed points is determined by the solution $h$ of the equation
\begin{equation}\label{eqh1}
h'(X)\left[C X  +F(X, h(X)\right[- \left[P h(X)+G(X,h(X))\right]=0
\end{equation}
In the case of $\mathcal{A}$  we have 
\begin{eqnarray}
&& C=0,\\
&& P= -\frac{3\lambda}{2\epsilon},\\
&& F=6 X^2-12 X^3,\\
&& G=\frac{3 M [2  (\lambda +2 \epsilon )X (M-1)-\lambda  M]}{2 \epsilon },
\end{eqnarray}
and the \rf{eqh1} can be integrated exactly to give
\begin{equation}
h(X)=\frac{1-2 X}{X e^{\frac{\lambda }{4 X \epsilon }}-2 X+1}
\end{equation}
This function describes the Center Manifold for the non--hyperbolic fixed point $\mathcal{A}$. The same procedure can be used for the point $\mathcal{B}$ obtaining 
\begin{equation}
h(X)=-\frac{{X} e^{\frac{\lambda }{4 X\epsilon }}}{{X} e^{\frac{\lambda }{4 X\epsilon }}+2 {X}-1}
\end{equation}
In this way the phase space can be completely characterized.  Unexpectedly, it turns out that the only structural differentiation of the phase space is given by the ``orientation'' on the Saddele Nodes points ($\mathcal{A},\mathcal{B}$ whereas the others remain unchanged. The key parameter for the stability change is again $\epsilon$. Samples of its structure can be given in Figures \ref{PSExp01}  and \ref{PSExp02}.
 
\begin{table}[tbp] \centering
\caption{The fixed points and their stability of the purely fermion NMC model with $V(\varphi)=V_0 \exp(-\lambda \varphi)$. Here S= Saddle, A=attractor and SN=saddle node. The saddle nodes change orientation (but not character) with the sign of $\epsilon \lambda$.}
\begin{tabular}{cccccccc}
& & \\
\hline   Point &$(X,M,Y)$ & Eigenvalues &  Stability & \\ \hline
& & \\
$\mathcal{A}$ &$\left(-\frac{1}{2},0,-1\right)$&$\left[\frac{3\lambda}{2\epsilon}, 0\right]$ &  SN\\ 
& & \\
$\mathcal{B}$& $\left(-\frac{1}{2},1,0\right)$&$\left[-\frac{3\lambda}{2\epsilon}, 0\right]$& SN \\
& & \\
$\mathcal{C}$ &$\left(0,0,-1\right)$& $\left[-3 , -3 \right]$& S\\ 
& & \\
$\mathcal{D}$& $\left(0,1,0\right)$& $\left[-3 , 3 \right]$&S \\
& & \\ \hline
\end{tabular}\label{TableStabNOmEx}
\end{table}


\subsection{Dynamical Systems Analysis in presence of a perfect fluid.}
Let us now analyze the cosmology deriving from the action \rf{1.1} in presence of a perfect fluid. We are specifically interested in seeing how the presence of an additional perfect fluid affects the action of the non minimal coupling.  We choose therefore two different potentials: $V=V_0\exp(-\lambda\varphi)$  and $V=V_0(\varphi^2+V_1)^\gamma$. This choice is motivated by the fact that we want to explore potentials able to give rise to a cosmological term at late time. 

\subsubsection{The case $V=V_0\/exp(-\lambda \varphi)$}
As in the case without a perfect fluid, the characteristic function is given by  $\mathcal{V}=-\frac{\lambda  X}{(2 X+1) \epsilon }$. The general system (\ref{DynSys1}-\ref{constraint}) takes the form
\begin{eqnarray}
&&\frac{dX({\mathcal N})}{d{\mathcal N}}= -6 X^2-3 X,\label{DynSys1MV1}\\
&&\frac{dY({\mathcal N})}{d{\mathcal N}}= Y
   \left[M+(3 w+1) \Omega +\frac{ 3\lambda  X}{(2 X+1) \epsilon}+2\right]+Y^2 \left[2+\frac{ 3\lambda  X}{(2 X+1) \epsilon} \right],\label{DynSys2MV1}\\
&&\frac{dM({\mathcal N})}{d{\mathcal N}}= M^2+M \left[(3 w+1) \Omega +Y \left(2+
  \frac{3\lambda  X}{(2 X+1) \epsilon }\right)-1\right],\label{DynSys3MV1}\\
&& \frac{d\Omega({\mathcal N})}{d{\mathcal N}}=\Omega  \left[M-3 w+3 Y\frac{\lambda  X}{(2 X+1) \epsilon }+2 Y-1\right]+(3 w+1)\Omega^2 ,\label{DynSys4MV1}
\end{eqnarray}   
or implementing the constraint \rf{constraint} to eliminate $Y$
\begin{eqnarray}
&&\frac{dX({\mathcal N})}{d{\mathcal N}}= -6 X^2-3 X,\label{DynSys1MV1R}\\
&&\frac{dM({\mathcal N})}{d{\mathcal N}}= 3 M\left[\frac{\lambda  X (M+\Omega -1)}{\epsilon(2 X +1)}+M+(w+1) \Omega -1\right],\label{DynSys3MV1R}\\
&& \frac{d\Omega({\mathcal N})}{d{\mathcal N}}=3 \Omega  \left[\frac{\lambda  X (M+\Omega -1)}{\epsilon(2 X +1) }+M+(w+1) (\Omega -1)\right],\label{DynSys4MV1R}\\
 &&Y=\Omega+M-1.\label{constraint2}
\end{eqnarray}  
As in the $\Omega=0$ case, we perform a change of variables to work around the singularities of the system. Setting  again $d{\mathcal M}=d{\mathcal N} (1-2\epsilon \phi)$ we obtain
\begin{eqnarray}
&&\fl\frac{dX({\mathcal M})}{d{\mathcal M}}= -3 X(1+ 2X)^2\label{DynSys1MV1R},\\
&&\fl\frac{dM({\mathcal M})}{d{\mathcal M}}= \frac{3}{\epsilon} M\left[\lambda  X (M+\Omega -1)+\epsilon M (2 X +1)+\epsilon(w+1) (\Omega -1)(2 X +1)\right],\label{DynSys3MV1R}\\
&& \fl\frac{d\Omega({\mathcal M})}{d{\mathcal M}}=\frac{3}{\epsilon}  \Omega  \left[\lambda  X (M+\Omega -1)+\epsilon M (2 X +1)+\epsilon(w+1) (\Omega -1)(2 X +1)\right],\label{DynSys4MV1R}\\
 &&\fl Y=\Omega+M-1.\label{constraint3}
\end{eqnarray}
Both systems admit six fixed points and a one dimensional fixed subspace given in Tables \ref{TableMExp1}  and \ref{TableMExp2}  whose coordinates are independent of the parameters of the system. The set of fixed points includes the ones we have found in the $\Omega=0$ case plus two new fixed points which represent states in which the matter contributions are dominant. One of these points ($\mathcal{F}$) is associated with standard Friedmann cosmologies. The other ($\mathcal{C}$) corresponds to an exponential in the case of dust ($w=0$) and a power law otherwise. This is an unusual behavior as a pressureless component is normally not expected to produce a de Sitter phase. However, since the condensate also acts as a form of dust (see the \rf{18.1} ) in the cosmology,  the theory does not ``recognize'' the presence of matter and shows a phenomenology typical of the $\Omega=0$ case. 
\begin{table}[tbp] \centering
\caption{The fixed points and the solutions of the NMC model with matter and $V(\varphi)=V_0 \exp(-\lambda \varphi)$.}
\begin{tabular}{cclllccc}
& & \\
\hline   Point &$(\Omega,X,M, Y)$ &  Scale Factor \\ \hline
& & \\
$\mathcal{A}$ &$\left(0,-\frac{1}{2},0,-1\right)$&  N/A \\ 
& & \\
$\mathcal{B}$& $\left(0,-\frac{1}{2},1,0\right)$& $a=a_0 \exp\left(\sqrt{-\frac{m}{12 \epsilon}}(t-t_0)\right)$ \\
& & \\
$\mathcal{C}$ &$\left(1,-\frac{1}{2},0,0\right)$&  \( a = \left\{ \begin{array}{ll}a_0\exp\left[\frac{\rho_0}{\varphi_0} (t-t_0)\right] & w=0\\a_0\left(t-t_0\right)^{2/3w}& w\neq0\end{array} \right. \) \\ 
& &\\
$\mathcal{D}$ &$\left(0,0,0,-1\right)$&  $a=a_0 \exp\left(\sqrt{-\frac{V_0}{6}}(t-t_0)\right)$\\ 
& & \\
$\mathcal{E}$& $\left(0,0,1,0\right)$& $a=a_0\left(t-t_0\right)^{2/3}$\\
& & \\
$\mathcal{F}$& $\left(1,0,0,0\right)$& $a=a_0\left(t-t_0\right)^{\frac{2}{3(1+w)}}$ \\
& & \\ \hline
& & \\
$\mathcal{L}$& $\left(\Omega_0,-\frac{1}{2},1-\Omega_0,0\right)$& \( a = \left\{ \begin{array}{ll}a_0\exp\left[\rho_0 \varphi_0 (t-t_0)\right] & w=0\\a_0\left(t-t_0\right)^{2/3w\Omega_0}& w\neq0\end{array} \right. \)  \\
& & \\ \hline
\end{tabular}\label{TableMExp1}
\end{table}

\begin{table}[tbp] \centering
\caption{The fixed points and the solutions of the NMC model with matter and $V(\varphi)=V_0 \exp(-\lambda \varphi)$.}
\begin{tabular}{cclllccc}
& & \\
\hline   Point &$(\Omega,X,M, Y)$ &  Condensate &Energy Density \\ \hline
& & \\
$\mathcal{A}$ &$\left(0,-\frac{1}{2},0,-1\right)$&  N/A&N/A \\ 
& & \\
$\mathcal{B}$& $\left(0,-\frac{1}{2},1,0\right)$&$\varphi\rightarrow\infty
$&$\rho=0$ \\
& & \\
$\mathcal{C}$ &$\left(1,-\frac{1}{2},0,0\right)$& $\varphi\rightarrow\infty
$& \( \rho = \left\{ \begin{array}{ll}0& w=0\\\rho_0(t-t_0)^{-2\frac{(1+w)}{w}}& w\neq0\end{array} \right. \)\\ 
& &\\
$\mathcal{D}$ &$\left(0,0,0,-1\right)$&$\varphi= 0$&$\rho=0$\\ 
& & \\
$\mathcal{E}$& $\left(0,0,1,0\right)$&$\varphi=0
$&$\rho=\rho_0(t-t_0)^{-2(1+w)}$ \\
& & \\
$\mathcal{F}$& $\left(1,0,0,0\right)$&$\varphi=0
$&$\rho=\rho_0(t-t_0)^{-2}$ \\
& & \\ \hline
& & \\
$\mathcal{L}$& $\left(\Omega_0,-\frac{1}{2},1-\Omega_0,0\right)$ &  $\varphi\rightarrow\infty
$& \( \rho = \left\{ \begin{array}{ll}0& w=0\\\rho_0(t-t_0)^{-2\frac{(1+w)}{w\Omega_0}}& w\neq0\end{array} \right. \) \\
& & \\ \hline
\end{tabular}\label{TableMExp2}
\end{table}

As usual, the stability has to be calculated using the HG theorem  for the hyperbolic points and the CMT for the non--hyperbolic ones. Only the stability of the non hyperbolic fixed points depends on the values of $\epsilon$. The results are shown in Table \ref{TableStabExp}.

Because of the higher dimensionality of the phase space the characteristic equation for the determination of the stability of the non hyperbolic fixed manifolds generalizes to:
\begin{equation}\label{eqh}
\mathbf{h}'(X)\left[C X  +F(X, \mathbf{h}(X))\right[- \left[\mathbf{P} \cdot \mathbf{h}(X)+\mathbf{G}(X,\mathbf{h}(X))\right]=0,
\end{equation}
where now $\mathbf{h}(X)=(h_1(X),h_2(X))$. Let us consider as an example the case of point $\mathcal{A}$.  In this case we 
have 
\begin{eqnarray}
&&\fl C=-24,\\
&&\fl \mathbf{P}=\left\{-6 (w+1) -\frac{3 \lambda }{2 \epsilon},-\frac{3 \lambda }{2 \epsilon}-6\right\},\\
&& \fl F= -6 \left\{(2 X+5) X^2+1\right\},\\
&&   \mathbf{G}=\left\{G_1, G_2\right\},\\
&& \fl G_1=\frac{3 \Omega}{2 \epsilon} \left\{\lambda  \left[2 X
   (M+\Omega-1)+M+\Omega\right]+4 \epsilon  \left[M
   (X+1)+(w+1) (\Omega X- X+\Omega)\right]\right\},\\
&& \fl G_2=\frac{3 }{2 \epsilon } \left\{\lambda \left[M^2-\Omega^2+2 M X
   (M+\Omega-1)\right]\right.\nn\\
&&\fl~~~~~~\left.+4 \epsilon 
   \left[M^2-\Omega^2+M X (M+w
   \Omega+\Omega-1)\right]\right\}.
   \end{eqnarray}   
The equations \rf{eqh} cannot be solved exactly. A standard approach in this case is a resolution by series. Using the approximation
\begin{equation}
h_1(X)= a X^2 +b X^3+...,\quad h_2(X)=d X^2 +e X^3+...,
\end{equation}
one solution for the leading coefficients is
\begin{eqnarray}
&&\fl a=-\frac{(\lambda +4 \epsilon ) \left\{\lambda ^2 (5 w-48)+16 [w (w^2-21w
   +83)+40] \epsilon ^2+16 \lambda  (w^2-16w+30) \epsilon \right\}}{32
   \epsilon  [\lambda +4 (w-7) \epsilon ][\lambda  (w-4)+2 (w^2-6w-8)
   \epsilon ]}, \\ && \fl d= \frac{\lambda ^2 (5 w-48)++16 [w (w^2-21w
   +83)+40] \epsilon ^2+16 \lambda  (w^2-16w+30) \epsilon }{32 \epsilon  [\lambda  (w-4)+2 (w^2-6w-8)
   \epsilon ]}.
\end{eqnarray}
In general the sign of these coefficients specifies the stability of the fixed points\footnote{It is worth to stress at this point that the solutions obtained in this way are not unique, but they can be proven to be topologically equivalent \cite{Car}.}. However, since the lower order term of the two expansions is even, the points represent a saddle node bifurcation and they are unstable.  For the points $\mathcal{B}$ and  $\mathcal{C}$ the determination of the stability is complicated by the fact that these points present two zero eigenvalues.   However only one of them is due to its non--hyperbolic character, while the other one represents  the fact that it belongs to the line $\mathcal{L}$. 

The entire line $\mathcal{L}$ lies on the invariant submanifold $X=-1/2$ and  therefore it is unstable for all the initial conditions with $X\neq-1/2$. In the invariant submanifold $X=-1/2$ its stability can be characterized by the projection of the system in the invariant submanifold itself. However, the submanifold $X=-1/2$ is irrelevant on the physical point of view and the above considerations combine with he ones of the $\Omega=0$ case are sufficient to characterize the line as unstable and points $\mathcal{B}$ and  $\mathcal{C}$  as saddle node bifurcations.

\begin{table}[tbp] \centering
\caption{The fixed points and their stability of the  NMC model  with matter and  $V(\varphi)=V_0 \exp(-\lambda \varphi)$. Here A=attractor, S=saddle point and SN= saddle node bifurcation (non hyperbolic).}
\begin{tabular}{cccccccc}
& & \\
\hline   Point &$(\Omega,X,M,Y)$ & Eigenvalues &  Stability \\  \hline
& & \\
$\mathcal{A}$ &$\left(0,-\frac{1}{2},0,-1\right)$&$\left[\frac{3\lambda}{2\epsilon},\frac{3\lambda}{2\epsilon}, 0\right]$ &  SN\\ 
& & \\
$\mathcal{B}$& $\left(0,-\frac{1}{2},1,0\right)$&$\left[-\frac{3\lambda}{2\epsilon},0, 0\right]$& SN\\
& & \\
$\mathcal{C}$ &$\left(1,-\frac{1}{2},0,0\right)$& $\left[-\frac{3\lambda}{2\epsilon},0, 0\right]$&SN  \\ 
& & \\
$\mathcal{D}$ &$\left(0,0,0,-1\right)$&$\left[-3 ,-3 -3(1+w) \right]$& A \\
& & \\ 
$\mathcal{E}$& $\left(0,0,1,0\right)$& $\left[-3 ,3, -3w \right]$& S \\
& & \\
$\mathcal{F}$& $\left(1,0,0,0\right)$&$\left[-3,3w, 3(1+w)\right]$&  S  \\
\\
& & \\ \hline
& & \\
$\mathcal{L}$&$\left(\Omega_0,-\frac{1}{2},1-\Omega_0,0\right)$& $\left[0,0, \frac{3\lambda}{2\epsilon}\right]$& SN  \\
& & \\ \hline
\end{tabular}\label{TableStabExp}
\end{table}

Since the phase space is three dimensional,  plotting the phase space and deducing geometrical information for it is not as easy as in the previous section. For this reason here and in the following subsection we will limit ourselves to give analytical considerations.
\subsubsection{The case $V=V_0 (\varphi^{2}+V_1)^\gamma$}
The potential $V=V_0 (\varphi^{2}+V_1)^\gamma$ is a simple extension of the power law potential considered in the previous section. It was chosen because of its relevance in inflationary scenarios in the framework of scalar field. At early times ($a\rightarrow0$) this potential coincides with the pure power law considered in the $\Omega=0$ case. At late time however ($a\rightarrow\infty$) it generates a cosmological term related to the value of the constant $V_1$.

In this case $\mathcal{V}=\frac{2\gamma X^2}{X^2+V_1  (2 X \epsilon +\epsilon )^2}$ and the general system (\ref{DynSys1}-\ref{constraint}) takes the form
\begin{eqnarray}
&&\fl \frac{dX({\mathcal N})}{d{\mathcal N}}= -3X (1+2X),\label{DynSys1MV1}\\
&&\fl\frac{dY({\mathcal N})}{d{\mathcal N}}= Y
   \left[M+(3 w+1) \Omega - \frac{6\gamma X^2}{X^2+V_1  (2 X \epsilon +\epsilon )^2}+2\right]\nn \\ &&+2Y^2 \left[1- \frac{ 3\gamma X^2}{X^2+V_1  (2 X \epsilon +\epsilon )^2}\right],\label{DynSys2MV1}\\
&&\fl\frac{dM({\mathcal N})}{d{\mathcal N}}= M^2+M \left\{(3 w+1) \Omega +2Y \left[1-
  \frac{ 3 \gamma X^2}{X^2+V_1  (2 X \epsilon +\epsilon )^2}\right]-1\right\},\label{DynSys3MV1}\\
&&\fl \frac{d\Omega({\mathcal N})}{d{\mathcal N}}=\Omega  \left\{M-3 w+2Y \left[1-
  \frac{ 3\gamma X^2}{X^2+V_1  (2 X \epsilon +\epsilon )^2}\right]-1\right\}+(3 w+1)\Omega^2 ,\label{DynSys4MV1}
\end{eqnarray}   
or implementing the constraint \rf{constraint} to eliminate $Y$
\begin{eqnarray}
&&\frac{dX({\mathcal N})}{d{\mathcal N}}=-3X (1+2X),\label{DynSys1MV1R}\\
&&\frac{dM({\mathcal N})}{d{\mathcal N}}= 3 M\left[M-\frac{2\gamma X^2 (M+\Omega -1)}{X^2+V_1  (2 X \epsilon +\epsilon )^2}+(w+1) \Omega -1\right],\label{DynSys3MV1R}\\
&& \frac{d\Omega({\mathcal N})}{d{\mathcal N}}=3 \Omega  \left[M-\frac{2\gamma X^2 (M+\Omega -1)}{X^2+V_1  (2 X \epsilon +\epsilon )^2}+(w+1) (\Omega -1)\right],\label{DynSys4MV1R}\\
 &&Y=\Omega+M-1.\label{constraint1}
\end{eqnarray}  
The fixed points  of the above system are shown in Table \ref{TableMV2}. In addition to the fixed points found in the previous section we find two more fixed points, whose coordinates are also independent of the parameters of the system. The character of the solutions associated with the fixed point $\mathcal{A}$ depends on the parameter $\gamma$. For $\gamma<1/6$, $\mathcal{A}$ is associated to decelerated expansion solutions whereas for  $1/6<\gamma<1/2$ it represents and accelerated expansion solution. The values $\gamma>1/2$ are associated instead to a contraction solution whose meaning is the same of the one given in the power law potential given in the previous section.  Apart $\mathcal{D}$, none of  the fixed points represents an exact solution for the cosmological equations, so that the behavior given in  Table \ref{TableMV2} represents again approximations of the general integral. The same caveats given in the previous sections hold here.  Compared to the power law potential examined in the $\Omega=0$ case, in this case  a second exponential solution appears as a late time solution as expected. However, upon substitution in the cosmological equations, this solution does not constitute a  de Sitter phase but rather an oscillatory solution.
\begin{table}[tbp] \centering
\caption{The fixed points and the solutions of the NMC model with matter and $V=V_0 (\varphi^{2}+V_1)^\gamma$.}
\begin{tabular}{cclllccc}
& & \\
\hline   Point &$(\Omega,X,M, Y)$ &  Scale Factor  \\ \hline
& & \\
$\mathcal{A}$ &$\left(0,-\frac{1}{2},0,-1\right)$&    $a=a_0\left(t-t_0\right)^{2/3(1-2\gamma)}$ \\ 
& & \\
$\mathcal{B}$& $\left(0,-\frac{1}{2},1,0\right)$& $a=a_0 \exp\left(\sqrt{-\frac{m}{12\epsilon }}(t-t_0)\right)$ \\
& & \\
$\mathcal{C}$ &$\left(1,-\frac{1}{2},0,0\right)$&  \( a = \left\{ \begin{array}{ll}a_0\exp\left[\frac{\rho_0}{\varphi_0}  (t-t_0)\right] & w=0\\a_0\left(t-t_0\right)^{2/3w}& w\neq0\end{array} \right. \) \\ 
& &\\
$\mathcal{D}$ &$\left(0,0,0,-1\right)$&  $a=a_0\exp\left(i\frac{|V_1|}{\sqrt{6}} (t-t_0)\right)$ \\ 
& & \\
$\mathcal{E}$& $\left(0,0,1,0\right)$& $a=a_0\left(t-t_0\right)^{2/3}$ \\
& & \\
$\mathcal{F}$& $\left(1,0,0,0\right)$& $a=a_0\left(t-t_0\right)^{\frac{2}{3(1+w)}}$ \\
& & \\ \hline
\end{tabular}\label{TableMV2_1}
\end{table}
\begin{table}[tbp] \centering
\caption{The fixed points and the solutions of the NMC model with matter and $V=V_0 (\varphi^{2}+V_1)^\gamma$.}
\begin{tabular}{cclllccc}
& & \\
\hline   Point &$(\Omega,X,M, Y)$ & Condensate &Energy Density \\ \hline
& & \\
$\mathcal{A}$ &$\left(0,-\frac{1}{2},0,-1\right)$ & $\varphi\rightarrow\infty $&$\rho=0$\\ 
& & \\
$\mathcal{B}$& $\left(0,-\frac{1}{2},1,0\right)$&$\varphi\rightarrow\infty$&$\rho=0$ \\
& & \\
$\mathcal{C}$ &$\left(1,-\frac{1}{2},0,0\right)$&  $\varphi\rightarrow\infty
$&\( \rho = \left\{ \begin{array}{ll}0& w=0\\\rho_0(t-t_0)^{-2\frac{(1+w)}{w}}& w\neq0\end{array} \right. \)\\ 
& &\\
$\mathcal{D}$ &$\left(0,0,0,-1\right)$& $\varphi=0$&$\rho=0$\\ 
& & \\
$\mathcal{E}$& $\left(0,0,1,0\right)$&$\varphi=0
$&$\rho=0$ \\
& & \\
$\mathcal{F}$& $\left(1,0,0,0\right)$&$\varphi=0
$&$\rho=\rho_0(t-t_0)^{-2}$ \\
& & \\ \hline
\end{tabular}\label{TableMV2}
\end{table}


Using the HG theorem one can find the stability of the fixed points which is given in Table \ref{TableStabMV2}.  The stability of the fixed points depends on all  the parameters of the system, including the nature of the perfect fluid considered. Of all the points only  $\mathcal{D}$ is an attractor for a large set of initial conditions and coincides effectively with a cosmology indistinguishable from a GR based one. The remaining points are always unstable so that the fact that they do not represent physical solutions is irrelevant for the  analysis of the cosmologies.

\begin{table}[tbp] \centering
\caption{The fixed points and their stability of the  NMC model  with matter and $V=V_0 (\varphi^{2}+V_1)^\gamma$. Here R= repeller, A=attractor and S=saddle point.}
\begin{tabular}{cccccc}
& & \\
\hline   Point &$(\Omega,X,M,Y)$ & Eigenvalues & $\gamma<1/2$& $1/2<\gamma<1$ & $\gamma>1$  \\ \hline
& & \\
$\mathcal{A}$ &$\left(0,-\frac{1}{2},0,-1\right)$&$\left[3,-3(1-2\gamma), -3(1+w-2\gamma)\right]$ &  S& R if  $ 0\leq w<2\gamma-1$  & R\\ 
& & &  S& S if  $ 2\gamma-1< w\leq 1$  & \\ 
& & \\
$\mathcal{B}$& $\left(0,-\frac{1}{2},1,0\right)$&$\left[3, -3w,3(1-2\gamma) \right]$& S & S &S \\
& & \\
$\mathcal{C}$ &$\left(1,-\frac{1}{2},0,0\right)$& $\left[3,3w,-3(1+w-2\gamma)\right]$& R& R if  $ 2\gamma-1 w\leq1$ &S \\ 
& & &  S& S if  $0\leq w<2\gamma-1$  & \\ 
& & \\
$\mathcal{D}$ &$\left(0,0,0,-1\right)$& $\left[-3, -3, -3 (1+w )\right]$&A & A & A\\
& & \\ 
$\mathcal{E}$& $\left(0,0,1,0\right)$& $\left[-3, 3, -3w\right]$& S & S &S \\
& & \\
$\mathcal{F}$& $\left(1,0,0,0\right)$& $\left[-3,-3w, 3(1+w)\right]$& S & S &S \\
& & \\ \hline
\end{tabular}\label{TableStabMV2}
\end{table}
Finally, also in this case, the dimensionality of the phase space prevents an effective graphical description of the phase space and we will not include them here.

\section{Conclusions}
In this paper we have analyzed the dynamic of the cosmology of a metric theory of gravity in which a condensate of fermions couples non-minimally to the geometry. One of the most interesting features of this model is the fact that the Dirac equations constrain the behavior of the condensate in a very strict way and this allows the use of the non--minimal coupling as a switch for additional terms in the action.  For example, consider the case of an action $\varphi^2 R+ \varphi R^2$. It is clear that since $\varphi\propto a^{-3}$ the Hilbert-Einstein term will be dominant at early time and the $R^2$ term will become important at late time. Instead an action $\varphi R+ \varphi^2 R^2$  will behave in the opposite way. In other words, using this feature of  the condensate one can regulate the influence of non--minimally coupled terms in a way which is impossible in the standard scalar tensor and higher--order gravity. In this sense we can speak of ``design'' of the actions in the case of theories with condensate NMC.

Our contribution in this framework is an analysis of the details of the cosmology via the Dynamical System Approach. The proposed formulation allows the exploration of a model which includes massive fermions, a perfect fluid and a completely general self--interaction potential for the condensate.  All these models present some common fixed points although the solutions for the scale factor associated with them can differ strikingly.  One of the most important results of our analysis is the confirmation of the possibility to control the behavior of the cosmology ``designing'' the action. In fact, the phase space analysis also highlights a somewhat expected feature: since the attractors of the theory are always a $\varphi=0$ state, this kind of theories always evolve towards states which are indistinguishable from GR. This result, added to careful design of the interaction potential like in the case of the exponential potential, can offer a natural dynamic way to generate  $\Lambda$(CDM) universes at late time.

Notwithstanding the general formulation of the DSA equations, the present work was specifically focused on two different types of potential: power law and exponential in absence and presence of a perfect fluid. The choice of these potentials is motivated by both the standard form of the potential considered in fermionic self interaction and the attempt to model an inflationary or dark energy phase.

In absence of a perfect fluid  the structure of the phase space makes clear that some interesting orbits  are present in the sector $-1/2<X<0$. In particular, for the case of a power law potential $V=V_0\varphi^\alpha$  and $\alpha<1$, the cosmology can start from an unstable de Sitter state ($\mathcal{B}$) and  evolve through a Friedmannian ($t^{2/3}$) behavior ($\mathcal{D}$) or a reduction of the expansion rate (via $\mathcal{A}$) to approach a power law evolution ($\mathcal{C}$). In particular for $0<\alpha<2/3$ the final power law phase of  $\mathcal{C}$ can be a power law inflation so that the orbits  $\mathcal{B}\rightarrow\mathcal{D}\rightarrow\mathcal{C}$ inludes an inflation era,  a transition to a Friedmannian cosmology and the onset of a (power law) dark energy era. This transition to dark energy domination was already found in \cite{Ribas} using a different approach. For $\alpha>1$, instead, the cosmology can start with a power law behavior  ($\mathcal{A}$)  and evolve towards  to a  Friedmannian  behavior ($\mathcal{D}$) either via an intermediate de Sitter  ($\mathcal{B}$)  or a power law decelerated expansion ($\mathcal{C}$). For $1<\alpha<5/3$  the power law phase of $\mathcal{A}$ corresponds to accelerated expansion (power law inflation). In this sense the orbits  $\mathcal{A}\rightarrow\mathcal{B}\rightarrow\mathcal{D}$ can describe the graceful exit from a (de Sitter) inflationary phase.

In the case of a exponential potential, one of the fixed points ($\mathcal{A}$) does not represent a solution for the system, but it is always unstable and does not constitute an issue for the understanding of the dynamics of the model. The orbits in the phase space of this case can contain up to two de Sitter phases and a Friedmannian one and one of the de Sitter phases ($\mathcal{C}$) is an attractor for every value of the parameters (although not for all the orbits/initial conditions). Particularly interesting are the orbits  in which the sequence de Sitter phase, Friedmann phase, de Sitter phase is realized. It is worth to stress again that from the properties of the behavior of the condensates the appearance of such behavior could be inferred already at action level.

As expected, the inclusion of a perfect fluid adds to the degrees of freedom of the cosmology and such change is reflected in an increase of the dimensionality of the phase space and the appearance of additional fixed points.  We have considered in this case the exponential potential  and a generalization of the power law potential considered in the previous cases.

In the case of an exponential potential in presence of a perfect fluid the picture that emerges is considerably more complicated than the pure fermionic one ($\Omega=0$). On top of the $\Omega=0$  fixed points other points appear together with a line of fixed points which are associated with power law and exponential solutions. The cosmology however still contains two de Sitter phases other than a proper Friedmann phase and it cannot be excluded that a sequence de Sitter phase, Friedmann phase, de Sitter phase is realized for a set of initial conditions of measure different from zero. 

In the case of a power law potential in presence of a perfect fluid an oscillating cosmology, in which phases of expansions are alternated to phases of contractions, becomes an attractor for a large set of initial conditions. The  period of the oscillations is strictly related to the constant $V_1$. The approaching trajectory can be, however, rather complicated and can be approximated with different types of power law behaviors and a de Sitter phase. It is worth stressing that this kind of behavior was not predictable {\it a priori} and this shows a limitation of the idea of the design of these models. We can construct a theory whose cosmology shows an exponential behavior but we cannot  easily control the character of the exponential solution.

An interesting phenomenon present in both the above cases is related to the fact that  since the condensate works as a dust fluid when these cosmologies are filled with dust one obtain behaviors typical of the $\Omega=0$ case (modulo the additional complication of the flow due to the increase of degrees of freedom). This happens because in the case of dust the theory is not able to ``distinguish'' between the perfect fluid and the condensate so that the $\Omega=0$  phenomenology appears again. This is clear from the structure of the solution associated with $\mathcal{C}$.

All in all, therefore, the presence of matter does have a deep influence of the cosmological models, but in some cases, like the one of the exponential potential, it offers the possibility of a full representation of an inflation plus $\Lambda$ universes. 

The analysis of the above models  indicates that the possibility of a non--minimal coupling between a fermion condensate and the geometry can present, in spite of the additional constraint given by the Dirac theory, a phenomenology as rich as the standard scalar field theories. Although still at the toy model level of understanding, the theory presented above constitute an interesting alternative approach to the unification of inflation and dark energy which deserves further study.

To conclude it is worth stressing that one major issue for these models is to identify the fermion which would be able to generate the condensate considered above: at late time there is no obvious candidate for a fermion condensate. This is however not new in cosmology and particle physics. Probably the most important example is the theory of inflation in which, in spite of more that thirty years of study, the driving (scalar) field has never been found. In this sense, our work aims to highlight an alternative explanation for inflation and/or the cosmic acceleration phenomenon rather than prove the actual necessity of such mechanism.  Further studies will allow more critical review of these results.

\section{acknowledgments}
The authors would like to thank  Dr. L. Fabbri for useful discussions.


\end{document}